\documentclass[
reprint,
nofootinbib,
amsmath,amssymb,amsbsy,
prd,
onecolumn,
superscriptaddress,
longbibliography,
preprintnumbers]{revtex4-1}
\usepackage[dvipsnames,usenames]{xcolor}
\usepackage{graphicx}
\usepackage{graphicx}
\usepackage{mathtools}
\usepackage[section]{placeins}
\usepackage{dcolumn}
\usepackage{bm}
\usepackage{mathrsfs,natbib}
\usepackage[normalem]{ulem}
\usepackage[title]{appendix}
\usepackage{graphicx,epsf,amssymb,amsbsy,amsfonts,amssymb,amsmath}
\usepackage[colorlinks=true]{hyperref}
\hypersetup{
    unicode=false,          
    pdftoolbar=true,        
    pdfmenubar=true,        
    pdffitwindow=false,     
    pdfstartview={FitH},    
    pdftitle={},    
    pdfauthor={},     
    pdfsubject={},   
    pdfcreator={},   
    pdfproducer={}, 
    pdfkeywords={} {} {}, 
    pdfnewwindow=true,      
    colorlinks=true,       
    linkcolor=magenta, 
    citecolor=blue,        
    filecolor=magenta,      
    urlcolor=blue           
}
\usepackage{slashed}
\usepackage{comment}

\usepackage[dvipsnames]{xcolor}
\usepackage{color,soul}
\usepackage{simplewick}

\newcommand{\be}{\begin{equation}}
\newcommand{\ee}{\end{equation}}

\newcommand{\bfk}{{\mathbf k}}
\newcommand{\bfp}{{\mathbf p}}

\newcommand{\mybar}[1]

\newcommand{\beq}{\begin{eqnarray}}
\newcommand{\eeq}{\end{eqnarray}}

\newlength{\backup}

\begin{document}

\title{Chiral effects and Joule heating in hot and dense matter}
\author{Srimoyee Sen}%
\email{srimoyee08@gmail.com}
\affiliation{Department of Physics and Astronomy, Iowa State University, Ames, Iowa 50011, USA}%
\author{Varun Vaidya}%
\email{Varun.Vaidya@usd.edu}
\affiliation{Department of Physics, University of South Dakota, Vermillion, SD 57069, USA}%

\date{\today}

\begin{abstract}
Initial states of dense matter with nonzero electron chiral imbalance could potentially give rise to strong magnetic fields through chiral plasma instability. Previous work indicated that  unless chiral chemical potential is as large as the electron vector chemical potential, the growth of magnetic fields due to the instability is washed out by chirality flipping rate enabled by electron mass. We re-examine this claim in a broader range of parameters and find that at higher temperatures the hierarchy is reversed supporting a growing magnetic field for an initial electron chiral chemical potential much smaller than the electron vector chemical potential.  
Further, we identify a qualitatively new effect relevant for magnetized hot and dense medium where chiral magnetic effect (CME) sourced by density fluctuation acts as a powerful source of Joule heating. Remarkably, even modest chiral chemical potentials (keV) in such environment can deposit energy densities set by the QCD scale in a relatively short time of the order of a few milliseconds or seconds. We speculate how this mechanism makes CME-driven Joule heating a potentially critical ingredient in the dynamics of turbulent density fluctuation of supernovae and neutron star mergers.
\end{abstract}
\maketitle

\section{Introduction}
Massless chiral fermions are interesting because of their role in the chiral anomaly which relates the non-conservation of axial charge with the topological charge stored in gauge fields. Physical consequences of this interplay have enjoyed sustained interest over the past two decades especially in the study of hot and dense matter like the quark gluon plasma \cite{Fukushima:2008xe,Tuchin:2025bll,Manuel:2015zpa,Tuchin:2014iua, Tuchin:2019jxd}, early universe physics as well as cold dense matter in neutron star physics \cite{Akamatsu:2013pjd, Kharzeev:2013ffa, Kaplan:2016drz, Grabowska:2014efa,Yamamoto:2015gzz, Yamamoto:2022yva, Huang:2017pqe, Ohnishi:2014uea, Sen:2016jzl}. The associated phenomena include chiral magnetic effect \cite{Fukushima:2008xe}, chiral vortical effect \cite{PhysRevD.20.1807, Kharzeev:2007tn, PhysRevLett.103.191601,Landsteiner:2011cp}, chiral charge separation, chiral plasma instability \cite{Akamatsu:2013pjd,Kaplan:2016drz, Grabowska:2014efa, Sen:2016jzl}, chiral magnetic and vortical waves \cite{Kharzeev:2010gd, Jiang:2015cva} among others. The focus of this paper is the chiral magnetic effect (CME) and associated phenomenology for dense matter physics, e.g. in proto-neutron stars, merging neutron star (NS) environments. To a first approximation, these environments can be modeled as a plasma of degenerate neutrons and electrons with degenerate or non-degenerate protons.  We discuss two main CME related phenomena relevant for such environments in this paper:
\begin{enumerate}
\item Chiral plasma instability (CPI) triggered by an electron chiral imbalance in dense matter,  
\item Joule/ohmic heating of dense matter with electron chiral imbalance in strong background magnetic fields.
\end{enumerate}

While the first of these two effects is widely studied in the literature, the latter is not. In this paper we re-examine the viability of the former in dense matter as well as propose the latter as a significant source of energy dissipation in hot and dense medium. These effects in their idealized form are observed in systems with massless chiral fermions coupled to gauge fields. For the NS environment, it is the electrons which one might consider to be light enough for such effects to arise. CME can be summarized as follows for idealized massless electrons with net chiral imbalance or chiral chemical potential $\mu_5$. A net $\mu_5$ implies that the populations of right and left chirality electrons, are unequal. This chiral imbalance modifies the Maxwell's equation inducing a current that is proportional to the chiral chemical potential and the magnetic field. The induced current is known as the CME current and has been derived using many different approaches for systems with and without a vector chemical potential \cite{Fukushima:2008xe,Son:2012wh, Kaplan:2016drz}. Maxwell's equation in the presence of the CME current can give rise to the two effects mentioned above, i.e. CPI and Joule heating, and we discuss them next.

{\bf Chiral plasma instability:} In the absence of any background electromagnetic(EM) field, Maxwell's equations with CME current can be unstable leading to the generation of strong electro-magnetic fields \cite{Akamatsu:2013pjd}. This instability is known as the chiral plasma instability. It was suggested in \cite{Ohnishi:2014uea} that CPI could explain the origin of strong magnetic fields observed in neutron stars. The proposal involved the setting of gravitational collapse of massive stars leading to the formation of proto-neutron stars during which rapid electron capture through weak interactions turns protons into neutrons. For idealized massless electrons, this electron capture via weak interactions would lead to an excess of right chirality electrons which can then induce the chiral plasma instability generating strong magnetic fields. However, in reality electrons are not massless and the mass term can flip the chirality of right handed electrons into left thereby reducing the chiral imbalance and prevent the instability from taking place. It was first shown in \cite{Grabowska:2014efa} that at least in core collapse, the rate of chirality flip due to the mass term overcomes the rate of generation of chiral charge through weak interaction electron capture and damps out CPI. Since then, several other mechanism have been proposed which may be able to sustain sufficient chiral imbalance for long enough to give rise to strong magnetic fields \cite{Yamamoto:2022yva, Sigl:2015xva}. E.g. \cite{Sigl:2015xva} proposed that out of equilibrium electron plasma in the background of protons and neutrons may be able to sustain strong chiral imbalance. 

While discussing CPI in this paper, we don't concern ourselves with the mechanism of generating chiral imbalance and instead consider different initial states with a range of chiral chemical potential, temperature and electron vector chemical potential to analyze under what conditions initial states with chiral imbalance can generate strong magnetic fields. Analogous scenario were considered in \cite{Kaplan:2016drz, Dvornikov:2015iua} for cold degenerate matter in neutron stars. There \cite{Kaplan:2016drz} it was found that while an initial chiral imbalance for massless electrons will lead to CPI, the chirality flip rate for massive electrons is still too high for CPI to be effective. More precisely, it was found that the initial chiral chemical potential $\mu_5$ of the electrons must be as large as the electron vector chemical potential $\mu_e$ for the instability rate to be larger than the chirality flip rate. Only when such hierarchy of rates is achieved can the CPI generate strong fields. However, as pointed out in \cite{Kaplan:2016drz} itself, the regime of $\mu_5\sim \mu_e$ is on the border of the regime of validity of the calculation in \cite{Kaplan:2016drz}. In this paper we re-examine the validity of the constraint $\mu_5\sim \mu_e$ beyond the conditions of temperature and chemical potentials considered in \cite{Kaplan:2016drz, Dvornikov:2015iua, Grabowska:2014efa} and explore whether there exists any regime of parameters relevant for hot and dense matter where the CPI rate can be larger than the chirality flip rate while allowing for $\mu_5< \mu_e$. 

Note that, in dense matter with realistic electron mass, there are two impediments to CPI being effective in generating strong magnetic fields. The first is the chirality flipping due to the mass term that we have already discussed. The second relates to the extremely high electrical conductivity of dense media. In the absence of any conductivity, the CPI rate obtained from solving the Maxwell's equation is proportional to the chiral imbalance times the Fine structure constant, the rate being denoted as $\xi$, whereas in the presence of a nonzero electrical conductivity $\sigma$ the rate goes as $\sim (\sqrt{\xi^2+\sigma^2}-\sigma)$ which is much smaller than $\xi$ in the limit of $\sigma \gg \xi$. In a dense medium, especially with the hierarchy $\mu_e\gg \mu_5$, $\sigma$ is much larger than the scale $\xi$ which reduces the CME rate by orders of magnitude. One of our goals in this paper is to consider wider range of parameters to access regimes of $\mu_5, \mu_e$ and temperature $T$ that have not been considered before and to compare the CME rates to the chirality flip rate taking into account the variation of electrical conductivity in these regimes. We find that while for cold degenerate matter CPI rate is rather small, higher temperatures can cause CPI rates to be larger than the chirality flip rate even when $\mu_5< \mu_e$, e.g. $\mu_5\sim (\mathcal{O}(10)) \text{MeV}\ll \mu_e\sim \mathcal{O}(100)$ MeV. We consider degenerate electrons in the background of degenerate and non degenerate protons and find temperature enhancement of CPI in both regimes. While this enhancement in CPI allows for $\mu_5\ll \mu_e$, $\mu_5\sim 10$ MeV is still quite large and it is an open question as to whether such strong $\mu_5$ can be generated in neutron star environments.

{\bf Joule/ohmic heating:} This effect is different from the CPI even though it has the same origin as the CPI, i.e. the CME current. The effect can be understood by considering CME in the background of a strong magnetic field. Thus, we are not aiming to explain the generation of EM fields and instead quantifying the effects of the CME current in a pre-existing background EM field. In a medium with finite electrical conductivity, finite chiral chemical potential and a strong background magnetic field, the CME current will give rise to an electric field which will lead to dissipation through ohmic heating. Despite the large conductivity of the neutron stars, we find that such heating can be significant, depositing energy density set by the QCD scale, $\Lambda_{\text{QCD}}$, i.e. $\sim \Lambda_{\text{QCD}}^4$ to neutron star environments in a rather short time scale, of the order of a few milliseconds depending on the conditions as we discuss in the main text of the paper. This makes CME induced Joule heating particularly relevant for neutron star mergers and supernovae simulations which aim to understand the density and temperature profile of these objects \cite{Perego:2019adq, Fields:2023bhs, Radice:2017zta, Perego:2014fma, Radice:2020ddv}. The conditions needed for this type of heating are often realized in hot and dense matter: e.g. strong magnetic fields of the order of $10^{17}$ Gauss has been reported in merger simulations in \cite{Palenzuela:2021gdo, Kiuchi:2015sga}. Similarly, \cite{Sigl:2015xva} pointed out that chiral chemical potentials of the order of a few keV can be generated through the Urca process in the presence of density fluctuations. In this sense, the scenario of Joule heating we are considering here is somewhat different from the CPI setting we discussed above. In the CPI setting, discussed above, a chiral imbalance is assumed initially. But no chiral charge is pumped into the system subsequently. As a result $\mu_5$ is continuously depleted via chirality flipping due to electron mass. 
For the scenario of Joule heating, we are interested in a dense medium with density fluctuations such that the density fluctuation pumps chiral charge into the system continuously while the electron mass term flips chirality at the same time. The two opposing processes can realize a net background chiral chemical potential which can often reach a few keV \cite{Sigl:2015xva} and sustain over long time scales. The effect of this chiral imbalance on CPI itself was considered in \cite{Sigl:2015xva} and is not the focus of the current paper. Instead we focus on the quantification of Joule heating. 

The organization of this as paper is as follows. We begin with a quick review of the CPI and introduce the Joule heating mechanism. This is followed by an analysis of the viability of CPI taking into account chirality flipping due to electron mass in the regime of non-degenerate and subsequently, degenerate protons. We consider both $\mu_5\gg T$ and $T\gg \mu_5$ in our analysis. This is followed by a section on out of equilibrium physics in dense matter and  associated Joule heating estimate in supernovae and merger environments.

\section{Chiral magnetic effect}
In this section we will briefly review the physics of chiral plasma instability (CPI) and also discuss the dissipation due to CME in a background magnetic field. We begin with a degenerate Fermi gas of idealized massless electrons with a vector chemical potential $\mu_e$ and a net chiral imbalance which leads to a chiral chemical potential $\mu_5$. Note that, even for the highest temperatures considered in this paper, the electrons will remain highly degenerate. Since the electrons are relativistic, the temperature at which they become non-degenerate is set by their chemical potential. We will be using electron chemical potential $\mu_e$ of the order of $200$ MeV in this paper whereas the temperature in consideration do not exceed $T=60$ MeV. So, in formulas relating electron vector chemical potential to density, we ignore the temperature.
We can now write down the chemical potential for right and left chirality electrons as 
\beq
\mu_{\pm}=\mu_e\pm \frac{\mu_5}{2}.\nonumber
\eeq
If the density of right/left chirality electrons is denoted as $n_{\pm}$, we can write
\beq
n_{\pm}&=&\frac{\mu_{\pm}^3}{6\pi^2}\nonumber\\
n_e&\equiv& n_+ + n_-
\eeq
where $n_e$ is the net density of electrons and 
the chiral charge density $n_5$ being
\beq
n_5=n_+-n_-.
\eeq
Charge neutrality is ensured by maintaining a density of protons $n_p=n_e$. There are of course neutrons in a dense neutron star environment and we will consider the neutron density to be $n_n$.
It is well known that the Maxwell's equations get modified in this system with a CME current that was derived in \cite{Fukushima:2008xe, Grabowska:2014efa, Kaplan:2016drz} to be 
\beq
{\bf J}_{\text{CME}}=\xi \bf{B}
\eeq
with $\xi=\frac{\alpha_{\text{EM}}}{\pi}\mu_5$ where $\alpha_{\text{EM}}$ is the Fine structure constant. Maxwell's equation then modifies to
\beq
\bf\nabla \times \bf{B}-\frac{d\bf{E}}{dt}=\sigma \bf{E}+\xi \bf{B}.
\label{max}
\eeq 
\subsubsection{Chiral plasma instability}
It turns out with the CME term included in the Maxwell's equations as in Eq. \ref{max}, there is an instability in favor of growing electromagnetic fields. To see this effect 
we can directly solve for the instability by considering a gauge field ansatz in Coulomb gauge ( ${\bf\nabla} \cdot {\bf A} =0$)
\beq
A_0=0,\,\, {\bf{A}}=(\hat{x} \cos(k z)-\hat{y} \sin(k z))e^{t/\tau}A_k(0)
\label{ans}
\eeq
where $A_k(0)$ is the initial amplitude of the gauge field. Using $\bf{E}=-\frac{d\bf{A}}{dt}$ and $\bf{B}=\bf\nabla\times \bf{A}$ we can solve for the inverse growth rate $\tau$ of the growing mode as 
\beq
\tau=\frac{2}{\sqrt{4k(\xi-k)+\sigma^2}-\sigma}.
\eeq
The maximally growing mode corresponds to $k=\xi/2$ for which
\beq
\tau=\frac{2}{\sqrt{\xi^2+\sigma^2}-\sigma}.
\eeq
The corresponding growth rate is defined as 
\beq
\Gamma_{\text{CPI}}=\frac{1}{\tau}.
\eeq
This growth rate decreases with increasing electrical conductivity $\sigma$. The electric and magnetic field corresponding to Eq. \ref{ans} are helical and parallel to each other, i.e. 
\beq
{\bf{E}}_k=-\frac{\bf A}{\tau}, {\bf{B}}_k=k {\bf{A}}. 
\eeq
From this point on we will be focusing on the maximally growing mode, i.e. $k=\xi/2$ and eliminate subscripts from ${\bf{E}}_k$ and ${\bf{B}}_k$.
We see that $\bf{E}.\bf{B}\neq 0$ and as the EM fields grow, the topological charge stored in the gauge field, i.e. $\int d^3x \bf{E}.\bf{B}$ grows. If we denote the axial current of the electrons as $j_5^\mu$ where $n_5=j_5^0$, we know that the axial current suffers from the chiral anomaly \cite{PhysRevLett.42.1195}
\beq
\partial_\mu j_5^\mu=\frac{\alpha_{\text{EM}}}{4\pi} F\tilde{F}
\eeq
where $F$  is the electromagnetic field strength tensor and $\tilde{F}^{\mu\nu}=\epsilon^{\mu\nu\sigma\lambda}F_{\sigma\lambda}$ is the dual tensor. 
This implies that a growing $\bf{E}.\bf{B}$ will drain away chiral charge according to 
\beq
\frac{dn_5}{dt}=\frac{2\alpha_{\text{EM}}}{\pi}\int d^3x \,\,\bf{E}.\bf{B}.
\label{anom}
\eeq
 This would act as a back reaction and saturate the instability. In principle Eq. \ref{max} and Eq. \ref{anom} should be solved simultaneously to obtain the evolution of the electromagnetic (EM) modes. However, for initial stages of the instability when EM fields are small, it is reasonable to ignore the back-reaction. The growth rate obtained ignoring the back-reaction is of the same order of magnitude of the growth rate computed by solving the two equations Eq. \ref{max} and Eq. \ref{anom} simultaneously. We can now write the CPI rate in the limit of $\xi\ll \sigma$ as
\beq
\Gamma_{\text{CPI}}=\frac{\alpha_{\text{EM}}^2}{4\pi^2}\frac{\mu_5^2}{\sigma}.
\eeq 
In the next few sections we will consider CPI in several different regimes of parameters beginning with non-degenerate protons focusing on both $T\gg \mu_5$ and $T\ll \mu_5$. We will then move on to degenerate protons. Throughout this paper we will consider $\mu_e\gg T$ and $\mu_e\gg m_e$ such that electrons are degenerate and ultra-relativistic.
\subsubsection{Joule heating in a strong magnetic field}
Highly magnetized neutron star environments like magnetars and merging neutron stars can host very strong magnetic fields, sometimes of the order of $10^{18}$ Gauss $\sim (140\text{MeV})^2$ which is of the order of $\Lambda_{\text{QCD}}^2$ where $\Lambda_{\text{QCD}}$ is the QCD scale, assumed to be close to $\sim 100$ MeV here. We can consider the modified Maxwell's equation Eq. \ref{max} in this environment. The electromagnetic fields of interest in this case are macroscopic and therefore their spatial and temporal variation scale is much longer than the inverse mass scales corresponding to the micro-physics, i.e. set by the conductivity $\sigma$ and the chiral chemical potential $\alpha_{\text{EM}} \mu_5$. Thus, we can drop $\bf\nabla\times {\bf{B}}$ and $\frac{d\bf{E}}{dt}$ from the equation \ref{max} leading to 
\beq
\sigma {\bf E}=-\frac{\alpha_{\text{EM}}}{\pi}\mu_5 {\bf B}_0
\eeq
where ${\bf B}_0$ is the background magnetic field. This equation gives rise to an electric field proportional to the magnetic field given by
\beq
{\bf E}=-\frac{\alpha_{\text{EM}}}{\pi}\frac{\mu_5}{\sigma}{\bf B}_0.
\label{elec}
\eeq
This is evidently different from the chiral plasma instability which takes place when the initial electro-magnetic fields are small leading to growth of EM fields. Furthermore, the CPI generated electromagnetic fields with maximal growth rate have spatial and temporal dependence set by the the chiral chemical potential and the electrical conductivity as we discussed in the previous section. In the Joule heating scenario we are considering here, $\bf{B}_0$ and the electric field $\bf{E}$ in Eq. \ref{elec} do not require any space-time variation, e.g.  $\bf{B}_0$ is some background fields at par with a possible background chiral chemical potential $\mu_5$.

Given the electric field of Eq. \ref{elec} and a finite electrical conductivity, however large, we will expect power dissipation. The corresponding Joule heating rate is given by
\beq
J_E=\frac{1}{2}\sigma |{\bf E}|^2=\frac{\alpha_{\text{EM}}^2}{2\pi^2}\frac{\mu_5^2}{\sigma}|{\bf B}_0|^2.
\label{eq:JH}
\eeq
This rate of power dissipation holds for as long as $\mu_5$ or $\bf{B}_0$ survives. If the corresponding time scale is $\tau_B$, the dissipated energy is $\sim J_E \tau_B$. We will estimate the dissipated power for some realistic neutron star environments, in particular in the presence of density fluctuations in section \ref{Joule}, and find that the dissipated power can be significant. 
\section{CPI rates and chirality flipping due to electron mass}
In this section we will compute the chirality flipping rate due to electron mass in various conditions of temperature, electron chiral chemical potential, for degenerate and non-degenerate protons and compare it with the CPI rate to determine whether the instability can survive long enough to generate strong magnetic fields. In the process we will reproduce two sets of results: the chirality flip rate computed for non-degenerate protons for $T\gg \mu_5$ \cite{Grabowska:2014efa} and the chirality flip rate computed for degenerate protons for $T\gg \mu_5$ \cite{Dvornikov:2015iua} and used in \cite{Kaplan:2016drz}. The conclusions presented in \cite{Grabowska:2014efa, Dvornikov:2015iua, Kaplan:2016drz} involve further assumptions of hierarchy which we will discuss shortly. These additional hierarchies restrict to sub-regions of parameter space  where the chirality flip rates due to electron mass dominates over the CPI rate. In this paper we identify other hierarchies of parameters within the region of validity considered in \cite{Grabowska:2014efa, Dvornikov:2015iua, Kaplan:2016drz} where the CPI can survive and dominate over the chirality loss due to mass term. We also present new results for the mass-term induced chirality flip rate in the regimes of $\mu_5\gg T$ for both non-degenerate and degenerate protons. This leads to specific new insights into the regime $\mu_5^2\gg MT$, where $M$ is the proton mass. To elaborate, in general it is assumed that electron-electron scattering process is subdominant to electron-proton scattering process when computing the chirality flip rate due to electron mass. However, we find that for $\mu_5^2\gg MT$,  electron-electron scattering contribution dominates over that from electron-proton scattering. 
\subsection{CPI with non-degenerate protons}
\label{ndg}
We will now consider temperatures such that protons are non-degenerate, i.e. $T\gg T_p$ where $T_p$ is the proton degeneracy temperature set by $T_p=\frac{(\mu_P^2-M^2)}{2M}$ where $\mu_p$ is the proton Fermi energy and $M$ is the proton mass. In this regime, we can now consider initial chiral chemical potential of $\mu_5$ such that $\mu_e\gg \mu_5>0$ where $\mu_5$ can be larger or smaller than $T$. The limit of $T\gg \mu_5$ was considered in \cite{Grabowska:2014efa} whereas $T<\mu_5$ was not. We will consider both of these regions of parameter space in the following calculation.

We now compute the chirality flip rate of electrons induced by the electron mass. We will in general refer to this rate as $\Gamma_m$ and evaluate it indifferent parameter regimes. 
In an idealized scenario of massless electrons, this rate is zero. However, for massive electrons, electron chirality is not a good quantum number and as a result electron collision with electrons, protons, photon will lead to chirality flip. The dominant contribution to chirality flip in this regime arises from electron-proton scattering \cite{Grabowska:2014efa}. We also have another possible mechanism for chirality flipping, namely scattering of electrons with ambient photons, i.e. Compton scattering. The relevance of this process  for high temperature $T\gg \mu_e$ electron positron plasmas was pointed out in \cite{Boyarsky:2020ani}. However, in all our calculations, we assume that this channel is suppressed due to the low density ($\propto T^3$) of ambient photons since for all of the parameter space of interest $T \ll \mu_e$.

Since, chirality is not a good quantum number for massive electrons, we will use electron helicity as a proxy for chirality. 
E.g. We can begin with distribution functions for positive and negative helicity electrons assigning them chemical potentials $\mu_e^+, \mu_e^-$:
\beq
f_{\pm}({\bf k})=\frac{1}{1+e^{\beta(\sqrt{|{\bf k}|^2+m_e^2}-\mu_e^{\pm})}}\approx\frac{1}{1+e^{\beta(|{\bf k}|-\mu_e^{\pm})}},
\label{orig}
\eeq
where $m_e$ is the electron mass and the last approximate sign holds as long as $m_e\ll \mu_e$. 
The helicity imbalance in density can be expressed as
\beq
n_{\text{h}}=\int \frac{d^3 k}{(2\pi)^3}(f_+ - f_-)
\eeq
whereas the chirality imbalance in density is expressed as 
\beq
n_{\text{5}}=\int \frac{d^3 k}{(2\pi)^3}(f_+ - f_-)\frac{|{\bf k}|}{\sqrt{|{\bf k}|^2+m_e^2}}
\eeq
 as long as $\mu_e^{\pm}\gg m_e$, $n_{\text{h}}\approx n_5+ O(m/\mu_e)$. Thus, in the limit of $m\ll \mu_e$, it is justified to substitute chirality flip rate with helicity flip rates \cite{Grabowska:2014efa}. Therefore, in the calculations below, we will not distinguish between the two and set $\mu_{\pm}=\mu_e^\pm$. We will denote the distribution function for the protons as $f_{\text{P}}$. Since proton mass is much larger than all other mass/energy scales in the problem, the chirality/helicity of protons is irrelevant to the discussion. 
$f_{\text{P}}$ is given by
\beq
f_{\text{P}}({\bf k})=\frac{1}{e^{\beta(\sqrt{|{\bf k}|^2+M^2}-\mu_\text{P})}+1}
\eeq
where $\mu_\text{P}$ is the proton chemical potential  $\mu_\text{P}=\sqrt{k_F^2+M^2}$. 

The Boltzmann eqn. for $f_+$ capturing chirality flip due to electron mass reads
\beq
\dot{f}_+({\mathbf k})&&=\int \frac{d^3{\mathbf k}'}{(2\pi)^9}\frac{d^3{\mathbf p}'d^3{\mathbf p}}{2\omega_{{\mathbf k}}2\omega_{{\mathbf k}'}2\omega_{{\mathbf p}}2\omega_{{\mathbf p}'}} (2\pi)^4\delta^4(p'+k'-p-k)\nonumber\\
&&\Big[|\mathcal{M}_{+-}|^2 f_-({\bf k'})(1-f_+({\bf k}))f_{\text{P}}({\bf p'})(1-f_{\text{P}}({\bf p}))-|{\mathcal{M}_{-+}|^2}f_+({\bf k})(1-f_-({\bf k'}))f_{\text{P}}({\bf p})(1-f_{\text{\bf P}}({\bf p'}))\Big]
\label{fdot1}
\eeq
where $\omega_{\bf k}$ and $\omega_{\bf k}'$ stand for electron energies with momenta ${\bf k}$ and ${\bf k}'$, $\omega_{\bf k}=k_0\sim|{\bf k}|$ and $\omega_{\bf k}'=k'_0\sim|{\bf k}'|$ respectively, whereas  $\omega_{\bf p}=p_0=\sqrt{|{\bf p}|^2+M^2}$ and $\omega_{\bf p}'=p'_0=\sqrt{|{\bf p}'|^2+M^2}$ stand for proton energies with momenta ${\bf p}$ and ${\bf p}'$.  $\mathcal{M}_{+-}$ is the amplitude for the negative helicity electrons going to positive helicity and $\mathcal{M}_{-+}$ is the amplitude for the reverse process. The corresponding expression for $\dot{f}_-$ is obtained by interchanging the $+$ and $-$ helicity symbols in Eq. \ref{fdot1}.
Manipulating the distribution function for the proton for momentum $\bf{p}$ , energy $p_0$ and momentum $\bf{p}'$ and energy $p_0'$ we get
\beq
f_{\text{P}}({\bf p})(1-f_{\text{P}}({\bf p'}))=\frac{1}{1+e^{\beta(p_0-\mu)}}\frac{e^{\beta(p_0'-\mu)}}{1+e^{\beta(p_0'-\mu)}}=f_{\text{P}}(p')(1-f_{\text{P}}(p)) e^{\beta(p_0'-p_0)}.
\eeq
Using $|{\mathcal{M}}_{+-}|^2=|{\mathcal{M}}_{-+}|^2$  and energy conservation, the expressions further simplify to
\beq
\dot{f}_+({\mathbf k})&=&\int \frac{d^3{\mathbf k}'}{(2\pi)^9}\frac{d^3{\mathbf p}'d^3{\mathbf p}}{2\omega_{{\mathbf k}}2\omega_{{\mathbf k}'}2\omega_{{\mathbf p}}2\omega_{{\mathbf p}'}}
{|\mathcal{M}_{+-}|^2}(2\pi)^4\delta^4(p'+k'-p-k)\times\nonumber\\
&&\hspace{1.5in}\left(\frac{e^{\beta(|{\bf k}|-\mu_e^+)}-e^{\beta(|{\bf k}|-\mu_e^-)}}{(e^{\beta(|{\bf k}|-\mu_e^+)}+1)(e^{\beta(|{\bf k}'|-\mu_e^-)}+1)}\right)f_{\text{P}}({\bf p'})(1-f_{\text{P}}({\bf p}))
\label{fpdot}
\eeq
and 
\beq
\dot{f}_-({\mathbf k})&=&\int \frac{d^3{\mathbf k}'}{(2\pi)^9}\frac{d^3{\mathbf p}'d^3{\mathbf p}}{2\omega_{{\mathbf k}}2\omega_{{\mathbf k}'}2\omega_{{\mathbf p}}2\omega_{{\mathbf p}'}}
{|\mathcal{M}_{+-}|^2}(2\pi)^4\delta^4(p'+k'-p-k)\times\nonumber\\
&&\hspace{1.5in}\left(\frac{e^{\beta(|{\bf k}|-\mu_e^-)}-e^{\beta(|{\bf k}|-\mu_e^+)}}{(e^{\beta(|{\bf k}|-\mu_e^-)}+1)(e^{\beta(|{\bf k}'|-\mu_e^+)}+1)}\right)f_{\text{P}}({\bf p'})(1-f_{\text{P}}( {\bf p})).
\label{fmdot}
\eeq
\subsubsection{The regime of $T\gg\mu_5$}
\label{reg1}
Let's first consider the limit of $T\gg\mu_5$ in which case Eq. \ref{orig} leads to 
\beq
\dot{f}_+({\bf k})&\approx&\frac{e^{\beta(|{\bf k}|-\mu_e)}(\beta\dot{\mu}_5)}{2(1+e^{\beta(|{\bf k}|-\mu_e)})^2},\nonumber\\
\dot{f}_-({\bf k})&\approx&-\frac{e^{\beta(|{\bf k}|-\mu_e)}(\beta\dot{\mu}_5)}{2(1+e^{\beta(|{\bf k}|-\mu_e)})^2}
\label{fdots}
\eeq
where we have expanded in the small parameter $\mu_5/T$ and kept only the leading order term in the expansion. 
 Note that, even though generally, $\dot{f}_+({\bf k})\neq -\dot{f}_-({\bf k})$,  in the limit $T\ll \mu_5$, eq. \ref{orig} gives rise to $\dot{f}_+({\bf k})=-\dot{f}_-({\bf k})$ as seen in Eq. \ref{fdots} which then leads to
\beq
\dot{f}_+ -\dot{f}_- \approx\frac{e^{\beta(|{\bf k}|-\mu_e)}(\beta\dot{\mu}_5)}{(1+e^{\beta(|{\bf k}|-\mu_e)})^2}.
\label{fpmdot}
\eeq
We also expand the RHS of Eq \ref{fpdot} and \ref{fmdot} to linear order in $\mu_5/T$ and keep the leading order terms in this expansion which results in
\beq
\dot{f}_+({\mathbf k})&\approx&\int \frac{d^3{\mathbf k}'}{(2\pi)^9}\frac{d^3{\mathbf p}'d^3{\mathbf p}}{2\omega_{{\mathbf k}}2\omega_{{\mathbf k}'}2\omega_{{\mathbf p}}2\omega_{{\mathbf p}'}}
{|\mathcal{M}_{+-}|^2}(2\pi)^4\delta^4(p'+k'-p-k)\times\nonumber\\
&&\hspace{1.5in}\left(\frac{e^{\beta(|{\bf k}|-\mu_e)}(-\beta \mu_5)}{(e^{\beta(|{\bf k}|-\mu_e)}+1)(e^{\beta(|{\bf k}'|-\mu_e)}+1)}\right)f_{\text{P}}({\bf p'})(1-f_{\text{P}}({\bf p}))\nonumber\\
\label{lin1}
\eeq

and
\beq
\dot{f}_-({\mathbf k})&\approx&\int \frac{d^3{\mathbf k}'}{(2\pi)^9}\frac{d^3{\mathbf p}'d^3{\mathbf p}}{2\omega_{{\mathbf k}}2\omega_{{\mathbf k}'}2\omega_{{\mathbf p}}2\omega_{{\mathbf p}'}}
{|\mathcal{M}_{+-}|^2}(2\pi)^4\delta^4(p'+k'-p-k)\times\nonumber\\
&&\hspace{1.5in}\left(\frac{e^{\beta(|{\bf k}|-\mu_e)}(\beta\mu_5)}{(e^{\beta(|{\bf k}|-\mu_e)}+1)(e^{\beta(|{\bf k}'|-\mu_e)}+1)}\right)f_{\text{P}}({\bf p'})(1-f_{\text{P}}({\bf p})).\nonumber\\
\label{lin2}
\eeq

Combining Eq. \ref{fdots}, \ref{fpmdot}, \ref{lin1} and \ref{lin2} we can write:
\beq
\dot{\mu_5}=-2\mu_5 \int \frac{d^3{\mathbf k}'}{(2\pi)^9}\frac{d^3{\mathbf p}'d^3{\mathbf p}}{2\omega_{{\mathbf k}}2\omega_{{\mathbf k}'}2\omega_{{\mathbf p}}2\omega_{{\mathbf p}'}}
{|\mathcal{M}_{+-}|^2}(2\pi)^4\delta^4(p'+k'-p-k)\frac{(e^{\beta(|{\bf k}|-\mu_e)}+1)}{(e^{\beta(|{\bf k}'|-\mu_e)}+1)}f_{\text{P}}({\bf p'})(1-f_{\text{P}}({\bf p})).
\label{lin3}
\eeq
We now quote $\mathcal{M}_{+-}$ from \cite{Grabowska:2014efa, Dvornikov:2015iua}
 \beq
|\mathcal{M}_{+-}|^2=128\pi^2\alpha_{\text{EM}}^2\frac{m^2(1-\cos\theta)\omega_{\bf p}^2}{(2|{\bf k}|^2(1-\cos\theta)+m_D^2)^2}
\label{ruth}
\eeq 
where $m_D$ is the Debye scale and $\omega_{\bf p}$ is the energy of the proton assuming no recoil. Here, $\theta$ is the scattering angle for the electron $\theta =\theta_{\bfk \bfk'}$.
Computing the phase space integrals over ${\bf p}'$ and noting that the proton recoil is minimal, one can rewrite Eq. \ref{lin3} as 
\beq
\dot{\mu_5}=-2\mu_5 \int \frac{d^3{\mathbf k}'}{(2\pi)^6}\frac{d^3{\mathbf p}}{2\omega_{{\mathbf k}}2\omega_{{\mathbf k}'}4\omega_{\bf p}^2}
{|\mathcal{M}_{+-}|^2}(2\pi)\delta(|{\bf k}'|-|{\bf k}|)\frac{(e^{\beta(|{\bf k}|-\mu_e)}+1)}{(e^{\beta(|{\bf k}'|-\mu_e)}+1)}f_{\text{P}}({\bf p})(1-f_{\text{P}}({\bf p})).
\label{lin4}
\eeq
After computing the ${\bf p}$ and ${\bf k}'$ integrals one obtains the result for the flip rate for $k\sim \mu_e$ reproducing the results of \cite{Grabowska:2014efa}
\beq
\frac{\partial_t  \mu_5(\mu_e,t)}{\mu_5(\mu_e,t)}&=&\frac{\alpha_{\text{EM}}^2 m^2}{3\pi \mu_e}\left(\log\left(\frac{4\mu_e^2+m_D^2}{m_D^2}\right)-\frac{4\mu_e^2}{4\mu_e^2+m_D^2}\right)\nonumber\\
&\approx&
\frac{\alpha_{\text{EM}}^2 m^2}{3\pi \mu_e}\left(\log\left(\frac{4\mu_e^2}{m_D^2}\right)-1\right).
\label{gkr1}
\eeq
Thus, we find for this case 
\beq
\Gamma_m\approx\frac{\alpha_{\text{EM}}^2 m^2}{3\pi \mu_e}\left(\log\left(\frac{4\mu_e^2}{m_D^2}\right)-1\right).
\label{gkr2}
\eeq

\subsubsection{The regime of $T\ll\mu_5$}
\label{sub2}
We will now go to the opposite limit of $\mu_5\gg T$ which was not considered in \cite{Grabowska:2014efa}. In this limit, the linearization of the distribution functions in $\mu_5$ as shown in Eq. \ref{fdots} and \ref{fpmdot} does not apply. As a result, we will not attempt to compute 
$\partial_t\mu_5/\mu_5$.
However, we can compute the rate at which $f_+$ and $f_-$ change from the helicity flipping due to the mass term. More specifically, we will investigate $\frac{\dot{f}_+}{f_+}$ and $\frac{\dot{f}_-}{f_-}$ for states with $f_\pm\sim 1$. To compute $\dot{f}_+$, one can begin with the expression in Eq. \ref{fpdot} and perform the ${\bf p}'$ integral assuming minimal recoil leading to
\beq
\dot{f}_+({\mathbf k})&=&\int \frac{d^3{\mathbf k}'}{(2\pi)^6}\frac{d^3{\mathbf p}}{2\omega_{{\mathbf k}}2\omega_{{\mathbf k}}4 \omega_{\bf p}^2}
{|\mathcal{M}_{+-}|^2}(2\pi)\delta(|\bfk'|-|\bfk|)\left(\frac{e^{\beta(|{\bf k}|-\mu_e^+)}-e^{\beta(|{\bf k}|-\mu_e^-)}}{(e^{\beta(|{\bf k}|-\mu_e^+)}+1)(e^{\beta(|{\bf k}|-\mu_e^-)}+1)}\right)f_{\text{P}}({\bf p})(1-f_{\text{P}}({\bf p})).\nonumber\\
\label{fpdot2}
\eeq
In the above expression if we set $|{\bf k}|<\mu_e^-$ or $|{\bf k}|>\mu_e^+$ , $\dot{f}_+$ is exponentially suppressed in $e^{\beta(|{\bf k}|-\mu_e^-)}$, $e^{\beta(|{\bf k}|-\mu_e^+)}$ respectively. Therefore, it only makes sense to consider $\mu_e^+>|{\bf k}|>\mu_e^-$. Since, we are eventually interested in comparing the helicity flip rate due to the mass term with the CPI rate, we will consider $|{\bf k}|\sim \mu_e$ as a representative value, to obtain
\beq
\dot{f}_+&\approx&-\int \frac{d^3{\mathbf k}'}{(2\pi)^6}\frac{d^3{\mathbf p}'d^3{\mathbf p}}{2\omega_{{\mathbf k}}2\omega_{{\mathbf k}}4 \omega_{\bf p}^2}
{|\mathcal{M}_{+-}|^2}(2\pi)\delta(|\bfk'|-|\bfk|)f_{\text{P}}({\bf p})(1-f_{\text{P}}({\bf p}))\nonumber\\
&\approx&-\frac{\alpha_{\text{EM}}^2 m^2}{6\pi \mu_e}\left(\log\left(\frac{4\mu_e^2}{m_D^2}\right)-1\right)
\label{fpdot3}
\eeq
where we have used Eq. \ref{ruth}.
Similar arguments lead to an identical expression for $\dot{f}_-$, albeit with a positive sign. The rate in Eq. \ref{fpdot3} has the same parametric dependence as in Eq. \ref{gkr2} and interestingly, $\dot{f}_+-\dot{f}_-$ matches the rate in Eq. \ref{gkr2} exactly. Thus, if we take $\Gamma_m=|\dot{f}_+-\dot{f}_-|$ to stand in for the chirality flip rate for $T\ll\mu_5$, we see that the rate of chirality flip is insensitive to whether  $T\gg\mu_5$ or $\mu_5\gg T$.\\

\noindent
{\bf CPI rate compared to chirality flip rate with non-degenerate protons, for both $T\ll\mu_5$ and $T\gg \mu_5$:}
In order to obtain the CPI rate in this regime (non-degenerate protons with   $T\gg\mu_5$ or $\mu_5\gg T$) we need the conductivity in an environment with non-degenerate protons, which is given by
\beq
\sigma=\frac{\mu_e}{4\alpha_{\text{EM}}}\left(\log\left(\frac{4\mu_e^2}{m_D^2}\right)-1\right)^{-1}.
\label{eq:NDCond}
\eeq
This is a known result \cite{Grabowska:2014efa} which allows us to compute the ratio of the CPI rate $\Gamma_{\text{CPI}}$ to the chirality flip rate $\Gamma_m$ 
as
\beq
\frac{\Gamma_{\text{CPI}}}{\Gamma_m}=\frac{\alpha_{\text{EM}} \,\, \mu_5^2}{ m^2}\left(\frac{3}{\pi}\right)\sim .0278 \mu_5^2\,\, \text{MeV}^{-2}.
\label{ex1}
\eeq
From this expression it is clear that $\mu_5$ does not have to be as large as $\mu_e=200$ MeV for this ratio to be large. E.g. for $\mu_5\sim 20$ MeV, we obtain a ratio of $\sim 11$ enabling the instability\footnote{Note that, here, neither $\Gamma_{\text{CPI}}$ nor $\Gamma_m$ depend on temperature explicitly even though the protons are non-degenerate. This can be understood by noticing that the CPI rate only depends on the electron chiral imbalance and does not sense proton degeneracy. Furthermore, $\Gamma_m$ arising from electron scattering with non-degenerate protons can be expressed as $\sim n_\text{P} \sigma_R(\mu_e)$ \cite{Grabowska:2014efa} where $n_P$ is the proton density and $\sigma_R(\mu_e)$ is the Rutherford scattering cross-section for electrons at the Fermi surface. $\sigma_R(\mu_e)$ does not carry temperature dependence. Proton density $n_P$ on the other hand, using charge neutrality can be replaced by electron density which goes as $\mu_e^3$, thus eliminating any temperature dependence in the expression of $\Gamma_{\text{CPI}}$ and $\Gamma_m$ in this regime.}. 
\subsection{CPI with degenerate protons}
For degenerate protons, we have $T_p\gg T$. We can again consider two different regimes $T\ll \mu_5$ and $\mu_5\ll T$. Furthermore, the regime of $\mu_5\gg T$, can be further split into two regimes $MT\gg \mu_5^2$ and $MT\ll \mu_5^2$. The regime of $T\gg \mu_5$ was considered in \cite{Dvornikov:2015iua} and our calculation reproduces its results. Note that, although this regime differs from that of \ref{ndg}, several of its formulae will apply here as well.

Let's first consider $T \gg \mu_5$. In this limit, one can linearize the deviation in the distribution functions and compute $\frac{\dot\mu_5}{\mu_5}$ using Eq. \ref{lin3}. The new element is that $f_{\text{P}}$ now represents a degenerate distribution function. In the opposite limit ($\mu_5\gg T$), linearizing the distribution function in $\beta \mu_5$ is not valid and we will have to use $\dot{f}_+ - \dot{f}_-$ to quantify the flip rate. 
We compute it using Eq. \ref{fpdot} and \ref{fmdot}. Of course, one needs to consider the degenerate limit of $f_{\text{P}}$ when computing the ${\bf p}$ integral. 
Very similar to what we found in the case of non-degenerate protons, the  rate of $\frac{\dot\mu_5}{\mu_5}$ in the limit of $T\gg \mu_5$, and the rate $\dot{f}_+ - \dot{f}_-$ in the limit of $T\ll \mu_5$ end up coinciding  for $\mu_5^2\ll MT$. As a result, we will have a single expression for the flip rate that holds in these two regimes.  In other words, the expression for the rate of chirality flip rate induced by the electron mass, written in terms of the parameters of the theory, is insensitive to whether $\mu_5\gg T$ or $\mu_5\ll T$ as long as $\mu_5^2\ll MT$ holds for the latter.
For  $T\ll \mu_5$, $\mu_5^2\gg MT$, the situation changes as electron-electron scattering starts dominating the flip rate as we will see below.
\subsubsection{The regime of $T\gg\mu_5$}
\label{reg3}
In the limit of $\mu_5\ll T$, one can use the expression in Eq. \ref{lin4} with the expression for amplitude squared as given in Eq. \ref{ruth} treating the protons as a degenerate gas to write
\beq
\dot{\mu_5}=-2\mu_5 \int \frac{d^3{\mathbf k}'}{(2\pi)^6}\frac{d^3{\mathbf p}}{2\omega_{{\mathbf k}}2\omega_{{\mathbf k}'}4\omega_{\bf p}^2}
{128\pi^2\alpha_{\text{EM}}^2\frac{m^2(1-\cos\theta)\omega_{\bf p}^2}{(2|{\bf k}|^2(1-\cos\theta)+m_D^2)^2}}(2\pi)\delta(|{\bf k}'|-|{\bf k}|)\frac{(e^{\beta(|{\bf k}|-\mu_e)}+1)}{(e^{\beta(|{\bf k}'|-\mu_e)}+1)}f_{\text{P}}({\bf p})(1-f_{\text{P}}({\bf p})).\nonumber\\
\label{lin5}
\eeq
Using the result for $\int p^2 dp f({\bf p})(1-f({\bf p}))$ from Appendix \ref{app:DegP} we can write 
\beq
\dot{\mu}_5&=&-\frac{2\mu_5}{2\omega_{\bf k}}\int \frac{(k')^2 dk'}{(2\pi)^2}\frac{(d\cos\theta)}{2\omega_{\bf k'}}\frac{1+e^{\beta(\omega_{ \bf k}-\mu_e)}}{1+e^{\beta(\omega_{ \bf k'}-\mu_e)}}\left(32\pi\mu_P k_F T \right)\frac{\alpha_{\text{EM}}^2 m^2(1-\cos\theta)\delta(|\bf k|-|\bf k'|)}{(2|{\bf k}|^2(1-\cos\theta)+m_D^2)^2}.
\label{lin6}
\eeq
Substituting $|{\bf k}|\rightarrow\mu_e$, (using proton Fermi momentum $k_F\approx\mu_e$ and proton chemical potential $\mu_\text{P}=\sqrt{M^2+k_F^2}\approx M$ where $M$ is the proton mass)
\beq
\frac{\partial_t \mu_5(k,t)}{\delta \mu_5(k,t)}=-\frac{\alpha_{\text{EM}}^2}{\pi}\left(\frac{m}{\mu_e}\right)^2\frac{M}{\mu_e}T\left(\log\left(1+4\frac{\mu_e^2}{m_D^2}\right)-\frac{1}{1+\frac{m_D^2}{4\mu_e^2}}\right).
\eeq

This matches the result obtained in \cite{Dvornikov:2015iua} and at leading order in $\alpha_{\text{EM}}$, $\frac{m_D^2}{\mu_e^2}\sim 4\alpha_{\text{EM}}$ which results in the flip rate bring given by 
\beq
\Gamma_m\approx  \frac{\alpha_{\text{EM}}^2}{\pi}\frac{m^2}{\mu_e^3}MT \log\alpha_{\text{EM}}.
\label{flip2}
\eeq

For degenerate protons, the conductivity is \cite{Kaplan:2016drz, Baym1969}
\beq
\sigma=2\left(\frac{3}{\pi}\right)^{3/2}\frac{1}{3\pi^2 e}\frac{\mu_e^{9/2}}{M^{3/2}T^2}
\eeq
where $e=\sqrt{4\pi\alpha_{\text{EM}}}$. The ratio of CPI rate to chirality flip rate is then 
\beq
\frac{\Gamma_{\text{CPI}}}{\Gamma_m}\approx\frac{\pi^{5/2}}{4\sqrt{3}}\frac{e\sqrt{M}}{m^2 \mu_e^{3/2} }\frac{T \mu_5^2}{(-\log\alpha_{\text{EM}})}\equiv R
\label{R1}
\eeq
\subsubsection{The regime of $T\ll\mu_5$, $\mu_5^2\ll MT$}
\label{reg4}
In this regime, as before, we cannot linearize the distribution functions in $\beta\mu_5$. However, we can again study the behavior of $\dot{f}_+ -\dot{f}_-$ as enabled by the electron mass. Interestingly, the corresponding rate matches with Eq. \ref{flip2}. The calculational details are similar to those in section \ref{sub2} and we don't repeat them here. Thus the ratio of CPI rate to chirality flip rate due to the mass term is again given by Eq. \ref{R1}.

\subsubsection{The regime of $\mu_5\gg T$, $\mu_5^2\gg MT$}
\label{sub3}
We have thus far ignored electron-electron collision in the calculation of the chirality flip rate. As we will see this is valid as long as $\mu_5^2 \ll MT$. In the regime of $\mu_5^2> MT$ electron -electron collision can play a dominant role in chirality flip due to electron mass. To show this, we will compute $\dot{f}_\pm$ in this regime.   
Since we are considering electron-electron scattering, the recoiling particle is light and can change its momentum in contrast with electron-proton scattering. Note that, in the distribution functions below we will assume the electron to be massless since the electron mass is the subdominant effect here. It is only when we compute the scattering matrix element we will keep track of the nonzero electron mass since it gives the leading contribution to the flip amplitude. The corresponding Boltzmann equations are
\beq
&&\dot{f}_+({\mathbf k})\nonumber\\
&=&\sum_{h=\pm}\int \frac{d^3{\mathbf k}'}{(2\pi)^9}\frac{d^3{\mathbf p}'d^3{\mathbf p}}{2\omega_{{\mathbf k}}2\omega_{{\mathbf k}'}2\omega_{{\mathbf p}}2\omega_{{\mathbf p}'}}
{|\mathcal{M}_{+-}^{h}|^2}(2\pi)^4\delta^4(p'+k'-p-k)\left(\frac{e^{\beta(|{\bf k}|-\mu_e^+)}-e^{\beta(|{\bf k}|-\mu_e^-)}}{(e^{\beta(|{\bf k}|-\mu_e^+)}+1)(e^{\beta(|{\bf k}'|-\mu_e^-)}+1)}\right) f_h( {\bf p}')(1-f_h( {\bf p}))\nonumber\\
\label{fpdot1}
\eeq
and 
\beq
&&\dot{f}_-({\mathbf k})\nonumber\\
&=&\sum_{h=\pm}\int \frac{d^3{\mathbf k}'}{(2\pi)^9}\frac{d^3{\mathbf p}'d^3{\mathbf p}}{2\omega_{{\mathbf k}}2\omega_{{\mathbf k}'}2\omega_{{\mathbf p}}2\omega_{{\mathbf p}'}}
{|\mathcal{M}_{-+}^h|^2}(2\pi)^4\delta^4(p'+k'-p-k)\left(\frac{e^{\beta(|{\bf k}|-\mu_e^-)}-e^{\beta(|{\bf k}|-\mu_e^+)}}{(e^{\beta(|{\bf k}|-\mu_e^-)}+1)(e^{\beta(|{\bf k}'|-\mu_e^+)}+1)}\right) f_h( {\bf p}')(1-f_h( {\bf p}))\nonumber\\
\label{fmdot1}
\eeq
where $h$ the helicity of the recoiling particle. We are considering processes where the recoiling particle does not change its helicity in the scattering. $\mathcal{M}_{-+}^h$ is the amplitude for negative helicity electron going to positive helicity while the recoiling electron has helicity $h$. The definition of $\mathcal{M}_{+-}^h$ follows the same convention. We ignore helicity changing processes for the recoiling electron since they are further suppressed by powers of $m^2/\mu_e^2$.

Doing the ${\bf p}'$ integral taking $h=-$ for now(i.e. setting ${\bf p}'={\bf p+k-k'}\equiv \tilde{{\bf p}}$) we compute the negative helicity recoiling contribution to $\dot{f}_+$, which we denote as $\dot{f}_+^-$:
\beq
\dot{f}_+^- &=&\frac{1}{2\omega_{\bf k}}\int \frac{d^3{\bf k}'}{(2\pi)^32\omega_{{\bf k}'}}\left(\frac{e^{\beta(|{\bf k}|-\mu_e^+)}-e^{\beta(|{\bf k}|-\mu_e^-)}}{(e^{\beta(|{\bf k}|-\mu_e^+)}+1)(e^{\beta(|{\bf k}'|-\mu_e^-)}+1)}\right)\int \frac{d^3{\bf p}}{(2\pi)^34\omega_{{\bf p}}\omega_{{\bf p+k-k'}}}\left(|\mathcal{M}_{+-}^-\big|_{p'=\tilde{{\bf p}}}|^2\right)\nonumber\\
&&(2\pi)\delta(|{\bf p}|+|{\bf k}|-|{\bf k}'|-|{\bf p +k-k'}|)\left(\frac{1}{1+e^{\beta(|{\bf p +k-k'}|-\mu_e^-)}}\right)
\left(\frac{e^{\beta(|{\bf p}|-\mu_e^-)}}{1+e^{\beta(|{\bf p}|-\mu_e^-)}}\right).
\eeq
Without loss of generality, we can assume that $\bfk -\bfk'$ is along the z axis and $\bfp$ is in the x-z plane.  $\bfk, \bfk'$ must have the same azimuthal angle $\phi$ to ensure $\bfk -\bfk'$ is along the z axis. Hence their dot product will only depend on a single polar angle.
Let's consider the argument of the delta function where $\alpha$ is the angle between the incoming and outgoing helicity changing electron states ${\bf k}'$ and ${\bf k}$ and $\theta$ is the angle between $({\bf k}-{\bf k}')$ and ${\bf p}$. Thus, the delta function gives
\beq
&&|{\bf p}|+|{\bf k}|-|{\bf k}'|-|{\bf p +k-k'}|=0\nonumber\\
\implies && |{\bf p}|+|{\bf k}|-|{\bf k}'|-\sqrt{|{\bf p}|^2+|{\bf k}|^2+|{\bf k}'|^2-2|{\bf k}'||{\bf k}|\cos\alpha + 2|{\bf p}|\left(\sqrt{|{\bf k}|^2+|{\bf k'}|^2-2|{\bf k}||{\bf k'}|\cos\alpha}\right)\cos\theta}=0\nonumber\\
\implies && \cos\theta=\frac{-|{\bf k}||{\bf k}'|+|{\bf k}||{\bf p}|-|{\bf k}'||{\bf p}|+|{\bf k}||{\bf k}'|\cos\alpha}{|{\bf p}|\sqrt{|{\bf k}|^2+|{\bf k'}|^2-2|{\bf k}||{\bf k'}|\cos\alpha}}\equiv \cos\theta_0.
\label{betanot}
\eeq
Therefore, we can turn the delta function into
\beq
\delta(|{\bf p}|+|{\bf k}|-|{\bf k}'|-|{\bf p +k-k'}|)= \frac{\delta\left(\cos\theta-\frac{-|{\bf k}||{\bf k}'|+|{\bf k}||{\bf p}|-|{\bf k}'||{\bf p}|+|{\bf k}||{\bf k}'|\cos\alpha}{|{\bf p}|\sqrt{{\bf k}^2+{\bf k'}^2-2{\bf k}{\bf k'}\cos\alpha}}\right)}{\Big|-\frac{|{\bf p}|\sqrt{{|\bf k|}^2+|{\bf k'}|^2-2|{\bf k}||{\bf k'}|\cos\alpha}}{|{\bf p}|+|{\bf k}|-|{\bf k}'|}\Big|}.
\eeq
The azimuthal angle integrals for each momentum can be done trivially. Then writing $\frac{d^3{\bf p}}{(2\pi)^3}=\frac{|{\bf p}|^2 d|{\bf p}|\,\, d(\cos\theta)}{(2\pi)^2}$ and doing the $d(\cos\theta)$ integral we get
\beq
\dot{f}_+^-&=&\frac{1}{2\omega_{\bf k}}\int \frac{|{\bf k}'|^2}{(2\pi)^2}\frac{d(\cos\alpha)}{\omega_{|{\bf k}'}|}\frac{|{\bf p}|^2 d|{\bf p}|}{(2\pi)^2}\frac{|\mathcal{M}_{+-}^-|^2}{4\omega_{\bf p}(\omega_{\bf p}+\omega_{\bf k}-\omega_{{\bf k'}})}\left((2\pi)\frac{(|{\bf p}|+|{\bf k}|-|{\bf k}'|)}{|{\bf p}|\sqrt{|{\bf k}|^2+|{\bf k'}|^2-2|{\bf k}||{\bf k'}|\cos\alpha}}\right)\nonumber\\
&&\left(\frac{1}{1+e^{\beta(|{\bf p}|+|{\bf k}|-|{\bf k}'|-\mu_e^-)}}\right)
\left(\frac{e^{\beta(|{\bf p}|-\mu_e^-)}}{1+e^{\beta(|{\bf p}|-\mu_e^-)}}\right)\left(\frac{e^{\beta(|{\bf k}|-\mu_e^+)}-e^{\beta(|{\bf k}|-\mu_e^-)}}{(e^{\beta(|{\bf k}|-\mu_e^+)}+1)(e^{\beta(|{\bf k}'|-\mu_e^-)}+1)}\right).
\label{cosb}
\eeq

Let's take $|{\bf k}|\sim \mu_e$.
For small enough $T \ll \mu_e$, we can replace 
\beq
&&\left(\frac{1}{1+e^{\beta(|{\bf p}|+|{\bf k}|-|{\bf k}'|-\mu_e^-)}}\right)
\left(\frac{e^{\beta(|{\bf p}|-\mu_e^-)}}{1+e^{\beta(|{\bf p}|-\mu_e^-)}}\right)\left(\frac{e^{\beta(|{\bf k}|-\mu_e^+)}-e^{\beta(|{\bf k}|-\mu_e^-)}}{(e^{\beta(|{\bf k}|-\mu_e^+)}+1)(e^{\beta(|{\bf k}'|-\mu_e^-)}+1)}\right)\nonumber\\
&&\rightarrow -\theta(\mu_e^--|{\bf p}|)\theta(|{\bf k'}|-\mu_e^-)\theta(|{\bf p}|-\mu_e+\frac{\mu_5}{2})\theta(|{\bf p}|+\frac{\mu_5}{2}-|{\bf k'}|).
\eeq
From these constraints we conclude that the region of phase space that contributes in $\bfp$, $\bfk'$ plane is an isosceles right triangle with two sides each of length $\mu_5/2$, one side at $\bfp = \mu_e^-$ extending between $\mu_e^-+\mu_5/2>\bfk'\geq \mu_e^-$ and the other at $\bfk'=\mu_e^-$ extensing between $\mu_e^->\bfp\geq \mu_e^- -\mu_5/2$.  
Thus, the $|{\bf p}|$ integral and the $|{\bf k'}|$ integral can be replaced by $|{\bf p}|^2 d|{\bf p}|\,\, |{\bf k'}|^2 d|{\bf k'}|\rightarrow (\mu_e^-)^4 \mu_5^2/8\sim (\mu_e)^4 \mu_5^2/8$ while substituting $|{\bf p}|\rightarrow \mu_e^-\sim \mu_e$ and $|{\bf k'}|\rightarrow \mu_e^-\sim \mu_e$ in the rest of the integrand. To complete this calculation, we need the matrix elements $|\mathcal{M}_{+-}^h|^2$. 
We can write
 
\beq
\mathcal{M}_{+-}^-=e^2\bar{u}_-(k)\gamma^{\mu}u_+(k')\bar{u}_-(p)\gamma^{\nu}u_-(p')G_{\mu\nu}
\eeq
where $G$ is the relevant part of the photon propagator in medium given by
\beq
G_{\mu\nu}=\frac{g_{\mu\nu}}{(k-k')^2-m_D^2}
\eeq
and $u_-, u_+$ are electron spinors with negative and positive helicity.
 We don't need the general form of the corresponding amplitude squared for this calculation. Instead we only need  $|\mathcal{M}_{+-}^-|^2
\big|_{(p, p_0, k', k_0')\sim\mu_e^-, k=k_0\sim (\mu_e^-)}\equiv{\bf M}^2$. Using the definitions

\beq
u_-(k)&=&\begin{pmatrix}
-\sqrt{k_0+|{\bf k}|}e^{-i\phi}\sin\frac{\theta}{2}\\\\
\sqrt{k_0+|{\bf k}|}\cos\frac{\theta}{2}\\\\
-\frac{m}{\sqrt{k_0+|{\bf k}|}}e^{-i\phi}\sin\frac{\theta}{2}\\\\
\frac{m }{\sqrt{k_0+|{\bf k}|}}\cos\frac{\theta}{2}
\end{pmatrix}, u_+(k')=\begin{pmatrix}
\frac{m}{\sqrt{k_0'+|{\bf k}'|}}\cos\frac{\theta'}{2}\\\\
e^{i\phi'}\frac{m}{\sqrt{k_0'+|{\bf k}'|}}\sin\frac{\theta'}{2}\\\\
\sqrt{k_0'+|{\bf k}'|}\cos\frac{\theta'}{2}\\\\
e^{i\phi'}\sqrt{k_0'+|{\bf k}'|}\sin\frac{\theta'}{2}
\end{pmatrix},\nonumber\\
\nonumber\\\nonumber\\
u_-(p)&=&u_-(k)\big|_{|{\bf k}|\rightarrow |{\bf p}|, k_0\rightarrow p_0, \theta\rightarrow \tilde{\theta},\phi\rightarrow\tilde{\phi}}, \,\,\,\,
u_-(p')=u_-(k)\big|_{|{\bf k}|\rightarrow |{\bf p}'|, k_0\rightarrow p_0', \theta\rightarrow \tilde{\theta}',\phi\rightarrow\tilde{\phi}'},
\eeq
with our choice of co-ordinate system, $\phi'= \phi$ and $\theta- \theta' = \alpha$ we find at leading order in $\mu_5/\mu_e^-$ and $m/\mu_e^-$,

\beq
{\bf M}^2 = 16m^2\mu_e^2e^4 \frac{\cos^2\left(\frac{\theta_p-\theta'_p}{2}\right)\sin^2\left(\frac{\theta_k-\theta'_k}{2}\right) }{(2\mu_e^2(1-\cos(\theta_k-\theta'_k))+m_D^2)^2}= m^2\mu_e^2e^4 \frac{(1+\cos\left(\tilde \alpha\right))(1-\cos\left(\alpha\right))}{(\mu_e^2(1-\cos(\alpha))+m_D^2/2)^2}
\label{jc}
\eeq
where $\alpha$ is the angle between ${\bf k}$ and ${\bf k'}$ and $\tilde\alpha$ is the angle between ${\bf p}$ and ${\bf p'}$. In the evaluation of the phase space integrals, we computed an integral over $\cos(\theta)$ in Eq. \ref{cosb} where $\theta$ was the angle between ${\bf k}-{\bf k}'$ and $\bf p$. In order to make connection with the amplitude squared that we just computed in Eq. \ref{jc} we require the relation between $\theta$ and $\tilde\alpha$. This can be obtained using
\beq
{\bf p}'&=&{\bf k}+{\bf p}-{\bf k}'\nonumber\\
\implies{\bf p}.{\bf p}'&=&{\bf p}.({\bf k}+{\bf p}-{\bf k}')\nonumber\\
\implies\mu_e^2 \cos\tilde\alpha &\approx& 2\mu_e^2\cos\beta \sin\frac{\alpha}{2} + \mu_e^2
\label{angles}
\eeq
where in the last line we have substituted $|{\bf k}|\sim \mu_e,|{\bf k}'|\sim \mu_e^- \sim \mu_e, |{\bf k}'|\sim \mu_e^- \sim \mu_e $ assuming $\mu_5\ll \mu_e$. 
Using Eq. \ref{betanot} this gives
 $\cos \tilde \alpha = \cos \alpha$ which then leads us to 

\beq
{\bf M}^2\approx e^4 m^2\frac{(1-\cos^2\alpha)}{\mu_e^2\left((1-\cos\alpha)+\frac{m_D^2}{2\mu_e^2}\right)^2}.
\eeq
Therefore we can write $\dot{f}_+^-$ as
\beq
\dot{f}_+^-=\frac{e^4 m^2}{16\sqrt{2}(2\pi)^3}\frac{1}{\mu_e^3}\left(\frac{\mu_5}{2}\right)^2\int d(\cos\alpha) \frac{(1-\cos^2\alpha)}{(\sqrt{1-\cos\alpha})\left((1-\cos\alpha)+\frac{m_D^2}{2\mu_e^2}\right)^2}. 
\eeq
There is an identical contribution to $\dot{f}_+$ coming from scattering off of positive helicity electron. Completing a similar calculation for $\dot f_-$ we find 
\beq
\dot{f}_+ - \dot{f}_-=\frac{e^4 m^2}{8\sqrt{2}(2\pi)^3}\frac{1}{\mu_e^3}\left(\frac{\mu_5}{2}\right)^2\int d(\cos\alpha) \frac{(1-\cos^2\alpha)}{(\sqrt{1-\cos\alpha})\left((1-\cos\alpha)+\frac{m_D^2}{2\mu_e^2}\right)^2}. 
\eeq
We will ignore the cosine integral to get a parametric estimate of the chirality flip rate which is given by
\beq
\Gamma_m\sim\frac{\alpha_{\text{EM}}^2 m^2\mu_5^2}{8\sqrt{2}\pi\mu_e^3}.
\eeq
This is enhanced compared to the flip rate in Eq. \ref{flip2} by a factor of $\frac{\mu_5^2}{MT}$.
In this regime we have the ratio of CPI rate and flip rate as 
\beq
\frac{\Gamma_{\text{CPI}}}{\Gamma_m}\sim \frac{2\sqrt{2}\pi^3\sqrt{\alpha_\text{EM}} M^{3/2}T^2}{\sqrt{3} m^2 \mu_e^{3/2} } 
\label{ratl}
\eeq
For $\mu_e \sim 100$(s) MeV, taking $\mu_5 \sim 10$(s) MeV such that $\mu_5\ll \mu_e$, the constraint $\mu_5^2 >MT$ requires T to be very small, $< 0.1$ MeV, which makes this ratio of Eq. \ref{ratl} much smaller than $1$. Therefore we do not expect this regime to sustain a CPI.
\section{Summary of results for chiral plasma instability}

Our ultimate aim in this section is to review whether there is a regime in parameter space relevant for dense matter where the chiral instability persists and could  impact the dynamics of neutron stars. We will therefore use the analysis done so far to figure out the region of parameter space where the ratio of the CPI rate to the flip rate is greater than $1$. To illustrate this, we choose a typical value for the electron chemical potential in the core of neutron stars $\mu_e = 200$ MeV.
Having discounted regime $3$ discussed in section \ref{sub3} for degenerate protons, we focus on the analysis in regimes $1$ and $2$ for both degenerate and non-degenerate protons.\\ 

\noindent{\bf Non-degenerate protons:}
For non-degenerate protons, the ratio of CMI rate to chirality flip rate from electron mass term has a simple expression given in Eq. \ref{ex1} which holds in both the parameter regime 1 ($T\gg \mu_5$) \ref{reg1} and 2  ($\mu_5\gg T$) \ref{sub2}. The expression clearly demonstrates that CPI rate can be much larger than the chirality flip due to mass term for a range of chiral chemical potential while maintaining $\mu_5\ll \mu_e$.\\

\begin{figure}[t!]
    \centering
    \includegraphics[width=0.5\linewidth]{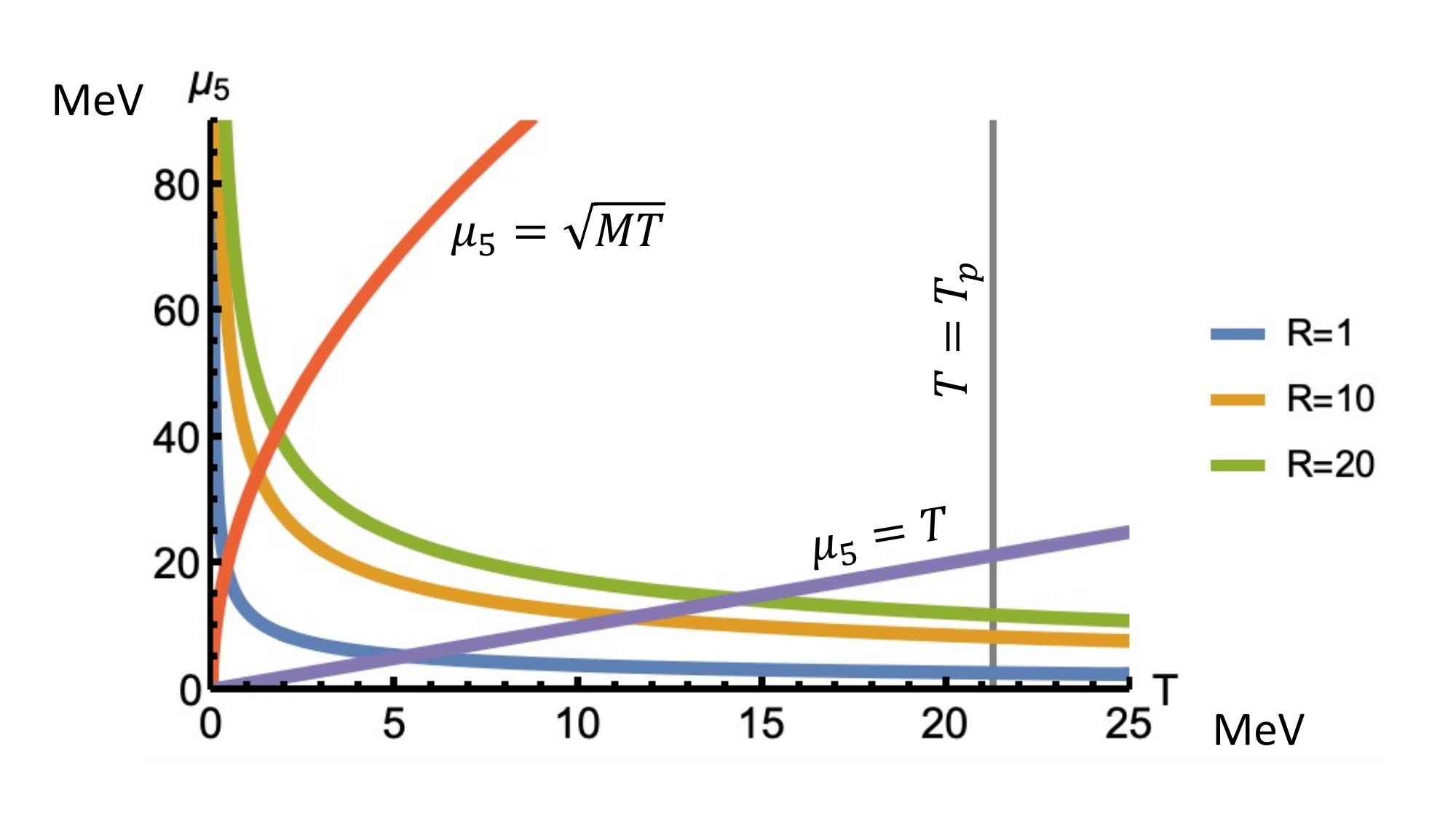}
    \caption{We plot $\mu_5$ as a function of temperature for which the ratio of the CPI rate to mass-term induced flip rate, $R$, given by Eq. \ref{R1} takes values equal to $1$, $10$, $20$. The expression of the flip rate in Eq \ref{R1} is valid within the range of the plot enclosed by the vertical red line and the purple line. The red vertical line separates the regime of parameter space for which $T<T_p$, i.e. degenerate protons and $T>T_p$ , i.e. non-degenerate protons. Clearly, the rate in Eq. \ref{R1} is valid for $T<T_p$. 
    The purple line separates the regime of $\mu_5>\sqrt{MT}$ and $\mu_5<\sqrt{MT}$. The expression in Eq \ref{R1} is valid for $\mu_5<\sqrt{MT}$. The orange line separates $\mu_5>T$ and $\mu_5<T$. The flip rate given by Eq. \ref{R1} is valid for both $\mu_5\gg T$ and $T\gg \mu_5$. We push the flip rate to the border of its region of validity $\mu_5\sim T$. }
    \label{param}
\end{figure}

\noindent{\bf Degenerate protons:}
Given that the ratio of the rates in regime 1 ($T\gg \mu_5$) \ref{reg3} and 2 ($\mu_5\gg T$) \ref{reg4} are identical, we use the common expression of Eq. \ref{R1} setting $M= 940$ MeV and $m=0.5$ MeV to plot the ratio of the rates in Fig. \ref{param}. We show the parameter space of $T$ and $\mu_5$ for which  R takes on specific values namely $R=1, 10, 20$. Note that although the expression in Eq. \ref{R1} holds both in the regime of $T\gg \mu_5$ as well as $\mu_5\gg T$, we don't have a closed form expression for this ratio in the intermediate regime $T\sim \mu_5$. In Fig. \ref{param} we push the expression of Eq. \ref{R1} to the border of the two parameter regions $1$ and $2$ for which it was derived.
It is quite clear that at lower temperatures $\mu_5$ has to be almost as large as $\mu_e$ or even larger for the instability to survive which is not phenomenologically viable except for extreme circumstances. This was the conclusion of \cite{Kaplan:2016drz}. However, for larger $T$, e.g. $T>1$ MeV, $\mu_5$ can be smaller than $\mu_e$ and still sustain the instability. \
Thus, we see that for degenerate protons, there are regions in the parameter space at relatively higher temperature where the instability rate is higher than the chirality flip rate due to the mass term. 

This implies that both non-degenerate ($T\gg T_p$) and degenerate protons ($T<T_p$) can sustain CPI and higher temperatures favor CPI growth.   
This is the main conclusion of  this paper with regards to the fate of CPI, that we wish to bring to the attention of the dense matter community. We further discuss the phenomenological applicability of this result in the conclusion. 
Note that, the exponential ansatz for the instability by itself does not tell us the value of the maximum magnetic field that can be reached via the instability. However, additional considerations of energy conservation as outlined in \cite{Kaplan:2016drz} shows that the instability saturates with a final magnetic field set by 
\beq
B=\frac{\mu_e\mu_5}{\sqrt{2}\pi} =(15\,\,\text{MeV})^2\left(\frac{\mu_5}{5\,\,\text{MeV}}\right)\left(\frac{\mu_e}{200\,\,\text{MeV}}\right)\sim 1.15 \times 10^{16}\left(\frac{\mu_5}{5\,\,\text{MeV}}\right)\left(\frac{\mu_e}{200\,\,\text{MeV}}\right)\text{Gauss}.
\eeq
This estimate corresponds to the maximally growing mode of $k=\xi/2$ which has a wavelength of inverse MeV scale. Thus, one would need other mechanisms like inverse cascade \cite{Hirono:2015rla, Gorbar:2016klv} to transfer the helicity to longer 
wavelengths which we don't discuss in this paper.
\section{Joule heating estimate}
\label{Joule}
In this section we return to the question of Joule heating in a strong magnetic field. We will estimate the Joule heating in a realistic scenario of hot neutron stars. In particular, we will consider the limit of non-degenerate protons and degenerate electrons as is the case in relatively hot dense environments of interest to us. We are interested in systems with density fluctuations which can occur in turbulent media as in merging NS or proto-neutron stars. At relatively higher temperatures, where protons are highly non-degenerate, the neutrinos may be trapped instead of free streaming. If the neutrino mean free path is larger than the density fluctuation length scales, the density fluctuations can lead to a build up of chiral chemical potential for electrons \cite{Sigl:2015xva}. While chiral chemical potential can be generated this way, one may worry about the loss of chiral charge through chirality flipping of electron mass as we discuss in the CPI analysis. However, the analysis for a medium with density fluctuation where chiral charge is pumped in over an extended period of time is slightly different from the above CPI analysis. Here, we are not considering an initial chiral imbalance which then either converts to EM fields or gets washed out through chirality flipping. Instead, one can estimate a background chiral chemical potential that can be sustained over a long time by balancing the chirality flip rate due to electron mass with the chiral charge generation rate due to density fluctuation. This is to say we can write 
\beq
\frac{1}{n_5}\frac{dn_5}{dt}=-\Gamma_m + \frac{S_w}{n_5}
\label{eq}
\eeq
where $S_w$ is the rate of chiral charge generation through weak interaction process in density fluctuation. E.g. for core collapse, as shown in \cite{Grabowska:2014efa} $S_w=\Gamma_w n_e$ where $\Gamma_w$ is the electron capture rate and $n_e$ is the electron density. For hot and turbulent dense matter with trapped neutrinos, this rate was shown to be \cite{Sigl:2015xva} $\sim \Gamma_U T^3$ by Sigl and Leite where $\Gamma_U$ is the Urca electron capture rate and $T^3$ gives an estimate of the background neutrino population. Leite and Sigl \cite{Sigl:2015xva} computed this rate for several different temperatures scales for baryon density of about $n_B= 2n_0$ where $n_0$ is the nuclear saturation density. Using their rates for $S_w$, and setting the LHS of Eq. \ref{eq} to zero yields a background chiral chemical potential. According to Leite and Sigl \cite{Sigl:2015xva} a background $\mu_5$ of $10^{-3}$ MeV can be sustained from density fluctuations over length scales smaller than the neutrino mean free path at temperatures of about $40$ MeV. We now combine this with 
estimates for background magnetic field from the literature to compute possible Joule/ohmic heating in a magnetized medium. 
\cite{Palenzuela:2021gdo}, \cite{Kiuchi:2015sga} report strong magnetic fields of about $10^{16}-10^{17}$ Gauss observed in MHD simulations relevant for neutron star mergers. The corresponding time scales over which these magnetic fields are sustained are of the order of a few mili-seconds. For an estimate for the Joule heating, 
we assume the strongest magnetic field of about $10^{18}$ Gauss lasting for about $\tau_B$ of the order of $10^{-3} \text{s}\sim 10^{12} (\text{eV})^{-1}\sim 10^{18} (\text{MeV})^{-1}$. The CME current in such a magnetized medium in the presence of a background chiral chemical potential, will produce a Joule energy density (Eq.~\ref{eq:JH}) of
\beq
J_E=\frac{\alpha_{\text{EM}}^2}{2\pi^2}\frac{\mu_5^2}{\sigma}|{\bf B}_0|^2
\label{eq:JH2}
\eeq 
per unit time. 
We will estimate the energy density deposited over time $\tau_B$ for non-degenerate protons with $\sigma$ given by Eq.~\ref{eq:NDCond}, which allows us to write 

\beq
E_J&=& J_E \tau_B=\frac{2}{\pi^2}|{\bf B}_0|^2 \left(\alpha_{\text{EM}}^3\frac{\mu_5^2}{\mu_e}\right)\left(\ln \frac{4\mu_e^2}{m_D^2} -1\right)\tau_B\approx \frac{2}{\pi^2}|{\bf B}_0|^2 \left(\alpha_{\text{EM}}^3\frac{\mu_5^2}{\mu_e}\right)(-\ln\alpha_{\text{EM}})\tau_B \label{energy}.\eeq
If we substitute $\mu_5\sim 10^{-3}$ MeV and $\mu_e\sim 200$ MeV as estimated in \cite{Sigl:2015xva} for $T=40 $ MeV, in Eq. \ref{energy}, we obtain the following estimate for the Joule energy
\beq
E_J&&\approx 2\times 10^3 (140 \text{MeV})^4\left(\frac{|{\bf B}_0|}{10^{18}\text{Gauss}}\right)^2\left(\frac{\tau_B}{10^{-3}\text{s}}\right).
\label{rate}
\eeq
Considering an estimate for the QCD scale to be $\Lambda_{\text{QCD}}\sim 100$ MeV, we can rewrite the last line of Eq. \ref{rate} as 

\beq
E_J&\sim & 8\times 10^3 (\Lambda_{\text{QCD}})^4\left(\frac{|{\bf B}_0|}{10^{18}\text{Gauss}}\right)^2\left(\frac{\tau_B}{10^{-3}\text{s}}\right).
\label{rateb}
\eeq

For a temperature of $T\sim 20$ MeV, \cite{Sigl:2015xva} estimate $\mu_5\sim 10^{-6}$ MeV, for baryon density of the order of a few times nuclear saturation density. The corresponding Joule heating estimate from Eq. \ref{energy} for $\mu_e\sim 200$ MeV is given by
\beq
E_J
\sim  (30 \text{MeV})^4\left(\frac{|{\bf B}_0|}{10^{18}\text{Gauss}}\right)^2\left(\frac{\tau_B}{10^{-3}\text{s}}\right)
\label{rate2}\eeq
which in terms of the QCD scale can be expressed as 
\beq
E_J
\sim 0.008 (\Lambda_{\text{QCD}})^4\left(\frac{|{\bf B}_0|}{10^{18}\text{Gauss}}\right)^2\left(\frac{\tau_B}{10^{-3}\text{s}}\right).
\label{rate2}\eeq

From these estimates we see that the energy deposition is quite sensitive to the temperature, i.e. being much larger at higher temperatures. The rates obtained in Eq. \ref{rate} and Eq. \ref{rate2} are both large enough to have a significant effect on the dynamics of the medium. E.g. we would expect the Joule heating to raise the temperature of the medium which in turn will impact other transport coefficients. The heating will impact the Urca rates, neutrino trapping and as a consequence most likely the chiral chemical potential. In other words,
the effect of Joule heating through the generation of chiral chemical potential can feed back into the density fluctuation itself, potentially providing a new mechanism for the damping of density fluctuation. We leave a detailed analysis of these effects for future work.

\section{Discussion}
This paper carries a two fold message: the first relates to the fate of chiral plasma instability induced by initial state of electron chiral imbalance in dense matter relevant for neutron stars (NS), supernovae and NS mergers. The second relates to chiral magnetic effect in a strongly magnetized plasma which again is relevant for proto-neutron stars, magnetars, and more importantly for binary neutron star mergers. In the case of the former, i.e. the chiral plasma instability, we compute the chirality flipping rate due to electron mass over a wide range of parameter space and compare it to the CPI rate to answer whether CPI will be able to generate strong magnetic fields. The mass term induced flip rates computed for non-degenerate and degenerate protons from our calculation reproduce the results the literature when $T\gg \mu_5$,  whereas we also present flip rates for smaller temperatures, i.e. $T\ll \mu_5$, which were not considered previously. We find that, for sufficiently small temperatures $M T\ll \mu_5^2$, electron-electron scattering contributes the most to electron chirality flipping due to electron mass. Besides this, we also realize that there are exceptions to the previous findings which concluded that one needs electron chiral chemical potential to be almost as large as the
electron vector chemical potential, for the CPI rate to be larger than the chirality flip rate due to the mass term. In fact, we find a reverse hierarchy at higher temperatures, where the CPI rate can be larger than the flip rate due to the
mass term, for $\mu_5$ of the order $\mathcal{O}(10)$ MeV which is much smaller than the electron vector chemical potential $\sim 100-200$ MeV. It will be interesting to incorporate these results in neutron star merger simulations where sudden density increase can source chiral chemical potential through weak interaction processes. As an example, shock waves in NS mergers can raise the density of the downstream fluid by order one which can in principle source significant chiral charge. Whether such a scenario can generate MeV scale chiral chemical potential should be investigated. Finally, magnetohydrodynamics with chiral anomaly has been developed within the context of heavy ion collisions \cite{Hattori:2017usa}. It will be important to consider how this formalism can be extended to include neutron star environments. The formalism will involve including the CME current in MHD using constitutive relations for the chiral charge. 

Our second topic of interest is the Joule heating caused by the CME current from electron chiral imbalance generated by density fluctuation in a highly magnetized hot plasma. The conditions of interest to us could be found in merging neutron stars or highly magnetized proto neutron stars. Simulations of merging NS show temperatures that often reach a few MeV to several $100$ MeV \cite{Perego:2019adq}. Similarly, there is evidence of very strong magnetic fields, of the order of $10^{17}$ Gauss, being reached in some simulations \cite{Palenzuela:2021gdo, Kiuchi:2015sga}. Previous studies in the context of proto neutron stars \cite{Sigl:2015xva} show that density fluctuations in temperatures this high can result in a sustained electron chiral imbalance, which we propose can result in significant Joule heating in strong magnetic fields. We find that energy densities of the order of $\Lambda_{\text{QCD}}^4$ can be deposited in the system over a few seconds or milliseconds depending on the parameters.  
Note that, the analysis of \cite{Sigl:2015xva}
is performed in an environment with trapped neutrinos where density fluctuations occur over length scales smaller than the neutrino mean free path. However, analogous chiral chemical potential can also be generated if instead anti-neutrino trapping is favored as can be the case in certain merger scenarios \cite{Perego:2019adq}. Thus, it appears that Joule heating due to CME should be considered an essential ingredient of merger simulations of strongly magnetized medium. 
To take steps towards this goal, more detailed study of the Joule heating mechanism is needed that takes into account the feedback of the heating on transport coefficients themselves and how that back-reacts to the Joule heating itself. Similarly, a realistic neutrino distribution is needed to accurately model the electron chiral imbalance and its subsequent impact on the heating. Some recent studies have analyzed how strong magnetic fields can impact Urca rates \cite{Tambe:2024usx}. Strong magnetic fields are also likely to impact chirality flip rates. It is important to study their effects on Joule heating as well, which we leave for future work. 
\section{Acknowledgments}
We would like to thank Steven Harris and Sanjay Reddy for valuable comments on the manuscript. V.V. is supported by startup funds from the University of South Dakota and by the U.S. Department of Energy, EPSCoR program under contract No. DE-SC0025545. S.S. acknowledges support from the U.S.\ Department of Energy, Nuclear Physics Quantum Horizons program through the Early Career Award DE-SC0021892. We also thank Steven Harris for comments on the draft.  

\appendix
\section*{Appendix}

\section{}
\subsection{Non-degenerate proton distribution}
By non-degenerate protons, we mean the regime where $T \gg T_p $. We can then obtain a scaling for the proton chemical potential by demanding that the medium is electrically neutral. In this analysis we assume the electron is relativistic while the proton can be treated non-relativistically, i.e.$M \gg k_F \gg m_e $. Moreover we are  working in the regime where $T \ll \mu_e$.
This implies that 
\beq 
n_e=\frac{8\pi}{(2\pi)^3}\int p^2dp \frac{1}{1+e^{\beta (|\bfp| -\mu_e)}} \approx \frac{\mu_e^3}{3\pi^2}
\eeq
For the case of the proton with a chemical potential $\mu_{\text{P}}$, we can write 
\beq
n_{\text{P}}=\frac{8\pi}{(2\pi)^3}\int p^2 dp f_{\text{P}}(p) &=&\frac{8\pi}{(2\pi)^3}\int p^2dp\frac{1}{1+e^{\beta (\frac{|\bfp|^2}{2M}+M -\mu_P)}}\nonumber\\
&=& -\frac{1}{\pi^2}\frac{\sqrt{\frac{\pi}{2}}\text{PolyLog}\Big[\frac{3}{2}, -e^{\beta(\mu_P-M})\Big]}{\left(\frac{\beta}{M}\right)^{3/2}} \nonumber \\
&\approx & -\frac{1}{\pi^2}\frac{\sqrt{\frac{\pi}{2}}\text{PolyLog}\Big[\frac{3}{2}, -e^{\beta T_P}\Big]}{\left(\frac{\beta}{M}\right)^{3/2}} \approx \frac{1}{\pi^2\left(\frac{\beta}{M}\right)^{3/2}}
\eeq
Then the condition for charge neutrality yields 
\beq
\frac{1}{\left(\frac{\beta}{M}\right)^{3/2}}  =  \frac{\mu_e^3}{3} \implies 2MT \approx\mu_e^2
\eeq
Therefore we can compute 
\beq
\int p^2dp f_{\text{P}}(p)(1-f_{\text{P}}(p)) \approx \frac{1}{4} \left(\frac{2M}{\beta}\right)^{3/2} \approx \frac{\mu_e^3}{4} 
\eeq
\subsection{Degenerate proton distribution}
\label{app:DegP}
This is the regime where $T_p \gg T$. In this case, for the integral over the proton distribution function, we can write
\beq
\int p^2 dp f_{\text{P}}(p) =  -\frac{\sqrt{\frac{\pi}{2}}\text{PolyLog}\Big[\frac{3}{2}, -e^{\beta(\mu_P-M})\Big]}{\left(\frac{\beta}{M}\right)^{3/2}} \approx \frac{\sqrt{\frac{\pi}{2}}(\beta T_p)^{3/2}}{\left(\frac{\beta}{M}\right)^{3/2}\Gamma[3/2+1]} = \frac{(T_p M)^{3/2}}{3}.
\eeq
This follows from $\mu_P -M \approx T_p$ and utilizing the limit  $\text{PolyLog}[\nu, -e^{z}]_{z \rightarrow \infty} = -z^\nu/\Gamma[\nu+1]$.  
Again from charge neutrality and using $T_p= k_F^2/(2M)$ we can write 
\beq
k_F = \mu_e
\eeq
and can compute 
\beq
\int p^2dp f_{\text{P}}(p)(1-f_{\text{P}}(p)) \approx  \frac{M\mu_e}{\beta}.
\eeq
\bibliography{ref.bib}
\end{document}